\documentclass[fleqn,usenatbib]{mnras}
\usepackage[english]{babel}

\usepackage[T1]{fontenc}
\usepackage{ae,aecompl}

\usepackage{fancyhdr}
\usepackage{amsfonts}
\usepackage{amsmath}
\usepackage{amssymb}
\usepackage{multicol}
\usepackage{layout}
\usepackage{graphicx}
\usepackage{multirow}
\usepackage{color}
\usepackage{multirow}

\usepackage{times}
\usepackage{natbib}
\def\be{\begin{equation}}
\def\ee{\end{equation}}
\usepackage[outdir=./fig/]{epstopdf}

\newif\ifAMStwofonts
\AMStwofontstrue

\graphicspath{{./fig/}}

\title[Oscillating dark energy]{Observational constraints on the oscillating dark energy cosmologies}

\author[Mehdi Rezaei]{Mehdi Rezaei $^{1,2}$\thanks{rezaei@irimo.ir} \\
{$^1$ Research Institute for Astronomy and Astrophysics of Maragha (RIAAM)
, Maragha, Iran, P.O.Box:55134-441}\\
$^2$ Iran meteorological organization, Hamedan Research Center for Applied
Meteorology, Hamedan, Iran}

\date{Accepted ?, Received ?; in original form \today}

\pagerange{\pageref{firstpage}--\pageref{lastpage}} \pubyear{2019}

\begin{document}

\label{firstpage}

\maketitle

\begin{abstract}
In this study we combine the background and the growth rate data in order to study the ability
of the two oscillating dark energy parameterizations, to fit the observational data. Using the likelihood and MCMC method we try to explore the posterior space and put constraints on the free parameters of the models. Based on the values of the well known Akaike and Bayesian information criteria we find that both of oscillating
dark energy models considered in this work are disfavored by the combined (background+growth rate) data. Although using the expansion data we can not reject oscillating dark energy models, the combined analysis provides strong evidences against these models.
\end{abstract}

\begin{keywords}
 cosmology: methods: analytical - cosmology: theory - dark energy- large scale structure of Universe.
\end{keywords}

\section{Introduction}
A wealth of independent cosmological observations have confirmed the accelerated expansion of the Universe first inferred with observations of Type Ia supernova (SnIa)\citep{Riess1998,Perlmutter1999,Kowalski2008}. Some of these include cosmic microwave background (CMB)\citep{Komatsu2009,Jarosik:2010iu,Komatsu2011,Ade:2015yua}, baryonic acoustic oscillation (BAO)\citep{Tegmark:2003ud,Cole:2005sx,Eisenstein:2005su,Percival2010,Blake:2011rj,Reid:2012sw}, high redshift galaxies \citep{Alcaniz:2003qy}, high redshift galaxy clusters \citep{Wang1998,Allen:2004cd} and weak gravitational lensing \citep{Benjamin:2007ys,Amendola:2007rr,Fu:2007qq}. The understanding of this accelerated expansion, is one of the greatest challenges in physics.  Within the framework of General Relativity (GR), we must introduce a new form of fluid with sufficiently negative pressure,$p<-\rho/3$, that accounts for roughly $75\%$ of the total energy budget of the universe today. this new exotic component, generically dubbed dark energy (DE), is still unknown after more than a decade of theoretical and observational investigations. In its simplest form, the dark energy is assumed to be a cosmological constant or vacuum energy, with constant EoS parameter $w_{\rm \Lambda}=-1$ \citep{Peebles2003}. Although the $\Lambda$ cosmology is consistent with all the available observations\citep{Planck2013_XVI}, it suffers severe problems from a theoretical point of view, the fine-tuning and cosmic coincidence problems\citep{Weinberg1989,Sahni:1999gb,Carroll2001,Padmanabhan2003,Copeland:2006wr}. This
provides motivation to find new alternatives to overcome these problems, for example modifying the gravity or considering dynamical dark energy models with a time evolving energy density. These dynamical dark energy
models can be roughly classified into two groups, tracking models\citep{Steinhardt:1999nw} and scaling models\citep{Halliwell:1986ja}. One might consider non monotonicity in the dark energy EoS, $w_d(a)$ to solve the coincidence problem\citep{Linder:2005dw}. Thus, models with an oscillating EoS were introduced, because the present accelerated expansion phase would just be one of the many such phases occurring over cosmic history, especially at early times. Dodelson, Kaplinghat and Stewart gave a simple form of oscillating dark energy model which can provide a natural solution to the coincidence problem\citep{Dodelson:2001fq}. Nesseris and Perivolaropoulos have fitted several cosmological models using the Bayesian method and the SnIa data consisting of 194 data points\citep{Nesseris:2004wj}. Their best fits to the data have provided for an oscillating dark energy model among all the 13 ansatzes have been considered.  Xia,  Feng and Xhang have considered a simple two-parameter model of oscillating Quintom\citep{Xia:2004rw}. Their results indicated that a dynamical model of dark energy such as the oscillating Quintom they have considered, is mildly favored under SN, CMB and LSS data.

Feng and others have proposed a phenomenological model of oscillating Quintom which can alleviate the coincidence problem\citep{Feng:2004ff}. Their oscillating dark energy model accommodate the crossing of the phantom barrier, $w=-1$ as it is marginally suggested by observations\citep{Alam2004,Huterer2005,Choudhury:2003tj}. Riess and others have found the possibility of oscillating EoS by fitting a quartic polynomial of $w_d(z)$ to SnIa observations\citep{Riess:2006fw}

 Wei and Zhang have confronted various cosmological models with observational $H(z)$ data\citep{Wei:2006ut}. They have found that the best models have an oscillating feature for both $H(z)$ and $w_d(z)$, with the EoS crossing $-1$ around redshift $z\sim1.5$. Some other efforts have been done to investigate oscillating dark energy models which their results show that oscillating dark energy is consistent with observations\citep{Dutta:2008px,Liu:2009nv,Kurek:2008qt}. Some of these efforts lead to a better agreement with SnIa data when oscillating EoS dark energy is used instead of the cosmological constant or an EoS linearly dependent on the scale factor $a$\citep{Kurek:2007bu,Lazkoz:2010gz}. In the context of particle physics, it is possible to have an oscillating quintessence potential if one considers a pseudo-Nambu-Goldstone boson field when it has rolled through the minimum\citep{Dutta:2008px,Johnson:2008se}.An oscillating behavior can also be seen in models with growing neutrino mass, where the dark energy is coupled with massive neutrinos. In these models dark energy EoS oscillates at relatively low redshifts\citep{Amendola:2007yx,Baldi:2011es}. 
Therefore, oscillating dark energy as a kind of dynamical dark energy models not only can justify the accelerated expansion of the universe, but also can solve the coincidence problem. In order to study oscillating dark energy, one can examine non-monotonic potentials and investigate periodic behavior in EoS $w_d(a)$. In many cases such potentials do not give rise to a periodic EoS $w_d(a)$. For example the potential for a pseudo-Nambu-Goldstone boson field can be written as a clearly periodic form, $V(\phi)=V_0[ 1 + cos(\phi/f)]$, where $f$ is a (axion) symmetry energy scale\citep{Frieman:1995pm}. Nevertheless, unless the field has already rolled through its minimum, the EoS, $w_d(a)$ evolve in a monotonic form and indeed can be well described by the usual CPL parameterization, $w_d(a) = w_0 + w_1(1-a)$\citep{Linder:2005dw}. 
Therefore, in this paper following the method of Linder in \citep{Linder:2005dw}, we start directly with phenomenological periodic parameterizations for the EoS of dark energy. For such periodic parameterizations, there are some important parameters, the location of the center of the range about which $w_d(a)$ oscillates, the amplitude of oscillation, the frequency of oscillation and its phase. We investigate the effects of periodic behavior in $w_d(a)$ on cosmological observables. Here, for the first time, in addition to the geometrical data (including those of SnIa, CMB, BAO, BBN and Hubble expansion data), we use structure formation data ($f\sigma_8$) to study the growth of perturbations in two oscillating parameterizations for $w_d(a)$. Specifically, we are going to implement the likelihood analysis using the geometrical data to put constraints on the free parameters of our parameterizations. Then we repeat this analysis at the perturbation level using growth rate data. Finally, to complete our analysis, we perform an overall likelihood analysis including the geometrical + growth rate data to put constraints on the corresponding cosmological parameters and obtain best fit values of them. The outline of the paper is the following. In Sect.\ref{sect:parameterizations} we introduce two oscillating parameterizations for the EoS of DE. Afterward, in Sect.\ref{sect:BG} we present and discuss the background evolution of our oscillating DE models. Then we introduce our method and next we present our results at background level. We study the evolution of linear perturbation in oscillating DE cosmology in Sect. \ref{growth}. In this section we follow two steps. First attempt to put constraint on model parameters using growth rate data. Then add these data points on expansion data and perform an overall analysis. Finally we provide the main conclusions in Sect. \ref{conlusion}

\section{Oscillating parameterizations}\label{sect:parameterizations}

It is also well known that the EoS parameter plays an significant role in cosmology. The evolution of energy density of dark energy mainly depends on its EoS. Therefore, determining EoS of dark energy is one of the key tasks in cosmology. In this paper we focus on phenomenological oscillating parameterizations for the EoS of DE. Our first parameterization is

 \begin{eqnarray}\label{par1}
ODE1:    w_{\rm d}(a)=w_0- A\sin (B\ln a +\theta)\;.
\end{eqnarray}

where $w_0$ is the center of the range over which $w_d(a)$ oscillates. $A$ and $B$ are also amplitude and frequency of oscillation respectively and $\theta$ is the current value of the phase of the oscillation. Of course when the amplitude $A=0$ our parameterization reduced to a constant EoS. On the other hand, as the value of $A$ increases we likewise expect clear distinction from a model with constant EoS. The values we consider for the parameters $w_0$ and $A$, can specify the behavior of DE whether it evaluates in quintessence or phantom regime. In order to one can distinguish the oscillating DE from a model with constant EoS using cosmological observations, there are some limitations on the values of oscillation parameters. Reader can find this limitations in\citep{Linder:2005dw}. The other parameterization we consider in this work is meant to generalize the CPL parametrization \citep{Chevallier2001} in order to avoid the future unphysical divergence of the dark energy EoS typical of this model. In this parameterization, EoS has the form\citep{Ma:2011nc}

\begin{eqnarray}\label{par2}
ODE2:    w_{\rm d}(a)=w_0- A (a B\sin(\dfrac{1}{a}) +\theta)\;.
\end{eqnarray}  

It is easy to see that at distant future this new parameterization asymptotes to 

\begin{eqnarray}\label{par2infinity}
w_{\rm d}(a\rightarrow\infty)=w_0- A (B +\theta)\;.
\end{eqnarray}  
 Thus, considering this oscillating form for CPL parameterization, make it to avoid divergence at $a\rightarrow\infty$. In Eq.\ref{par2} same as previous parameterization (Eq.\ref{par1}), $w_0$, $A$, $B$ and $\theta$ are the central value of $w_d(a)$, amplitude, frequency and the phase shift of oscillation respectively. In the both of above parameterizations, the phase of the oscillation $\theta$, for simplicity is assumed to be zero.

\section{Background evolution in oscillating dark energy }\label{sect:BG}

In this section we study the background evolution in oscillating dark energy (ODE) cosmologies. Considering a spatially flat universe consists of radiation, non-relativistic matter and dark energy, the Hubble parameter $H\equiv {\dot a}/a$ takes the form

\begin{eqnarray}\label{frid1}
H^2=\frac{1}{3 M^2_{\rm p}}(\rho_{\rm r}+\rho_{\rm m}+\rho_{\rm de})\;,
\end{eqnarray}

where $M_{\rm p}$ is the reduced Plank mass and $\rho_{\rm r}$, $\rho_{\rm m}$ and $\rho_{\rm de}$ are the energy densities of radiation, dark matter and DE, respectively. Introducing the density parameter for fluid $i$ as $\Omega_{\rm i}= \frac{\rho_{\rm i}}{3 M^2_{\rm p} H^2}$ and replacing it in Eq.\ref{frid1} we can obtain Hubble parameter as

\begin{eqnarray}\label{fridf(a)}
H^2=\Omega_ {\rm r0} a^{-4}+\Omega_{\rm m0} a^{-3}+\Omega_{\rm d0} f(a)\;,
\end{eqnarray}

where $f(a)$ can be written as

\begin{eqnarray}\label{f(a)}
f(a)=\exp \left( -3\int_1^a \frac{1+w(a')}{a'}da' \right)\;.
\end{eqnarray}

Thus, replacing $w(a)$ from Eqs.(\ref{par1} \& \ref{par2}) in Eq.\ref{f(a)} and inserting its result and the current values of density parameters in Eq.\ref{fridf(a)}, one can obtain the evolution of Hubble parameter and the dimensionless Hubble parameter $E(a)=H(a)/H(a=1)$.
Bellow, we investigate the performance of oscillating parameterizations of EoS against the latest observational data. Specifically, we perform a statistical analysis using the expansion data including:

\begin{itemize}
\item 580 SnIa data from the Union2.1 sample \citep{Union2.1:2012}
\item The position of the CMB acoustic peak, $(l_a, R, z_\star)$, which given by $(302.40, 1.7246, 1090.88)$ from the WMAP data set \citep{Hinshaw:2012aka}. For more details we refer the reader to \citep{Mehrabi:2015hva}.
\item 6 data points from the BAO sample which includes distinct measurements of the baryon acoustic scale (see Tab. \ref{tabbao})
\item A data point for  Big Bang Nucleosynthesis (BBN) from \citep{Serra:2009yp,Burles:2000zk} 
\item 25 data points for $H(z)$ from the Hubble data (see Tab. \ref{tabh}).
\end{itemize}

Concerning the position of the CMB acoustic peak, it is more common to use the method of distance priors which are proposed as a compressed likelihood to substitute the full CMB power spectrum analysis \citep[see][]{Bond:1997wr,Efstathiou:1998xx,Wang:2007mza,Chen:2018dbv}. In these studies, CMB data are incorporated by using constraints on parameters $(R,l_a,\Omega_bh^2)$ instead of using the full CMB power spectra. It has been shown that measuring these parameters provide an efficient and intuitive summary of CMB data as far as dark energy constraints are concerned. Also in Ref.\citep{Chen:2018dbv}, the authors compared the distance prior method with the full CMB power spectra analysis by constraining some dark energy models and showed that the results of these methods are in full agreement. When the data points are correlated we can use the inverse of so-called covariance matrix describing the covariance between the data \citep{Verde:2009tu}. In the case of BAO sample, we know that these data points are not all independent, thus we use the inverse of covariance matrix, $C^{-1}_{BAO}$ which obtained by \citep{Hinshaw:2012aka} to solve the effects of dependent data points. More details can be find in \citep{Mehrabi:2015hva}. Moreover in order to make sure about independency of $H(z)$ data sets, We only include independent measurements of $H(z)$ from \citep{Farooq:2013hq}. In this paper authors provided a set of $28$ independent $H(z)$ measurements which $3$ points of them (\citep{Blake:2012pj} $H(z)$ points) are highly correlated with the $d_i$ points from \citep{Blake:2011en} in Table \ref{tabbao}. Therefore we pretermit these $3$ points and only use other $25$ data points as the Hubble data.

\begin{table}
 \centering
 \caption{BAO data set which we use in the current study.}
\begin{tabular}{c  c  c  c }
\hline \hline
 $z$  & $d_i$ & Survey & Reference \\
 \hline
  0.106 & 0.336  & 6df & \citep{Beutler:2011hx}  \\
 \hline
  0.44 & 0.0916 & WiggleZ & \citep{Blake:2011en}  \\
 \hline 
  0.6 & 0.0726 & WiggleZ &\citep{Blake:2011en} \\
  \hline
  0.73 & 0.0592 & WiggleZ & \citep{Blake:2011en} \\
 \hline
 0.35 & 0.113 & SDSS-DR7 &  \citep{Padmanabhan:2012hf}\\
 \hline 
  0.57 & 0.073 & SDSS-DR9 &  \citep{Anderson:2012sa} \\
  \hline \hline
\end{tabular}\label{tabbao}
\end{table}

\begin{table}
 \centering
 \caption{Hubble data set which we use in the current study. }
\begin{tabular}{c  c  c c }
\hline \hline
 $z$  & $H(z)$ & $\sigma_H$ & Reference\\
 \hline
0.070 & 69.0 &19.6 &  \citep{Zhang:2012mp} \\
 \hline 
 0.100 &69.0 & 12.0 & \citep{Simon:2004tf}\\
 \hline
 0.12 & 68.6 & 26.2  & \citep{Zhang:2012mp}\\
  \hline
 0.17&	83.0	& 8.0& \citep{Simon:2004tf}\\
  \hline
0.179&	75.0&	4.0& \citep{Moresco:2012jh}\\
  \hline
0.199&	75.0	&5.0 & \citep{Moresco:2012jh}\\
  \hline
0.2	&72.9	& 29.6 & \citep{Zhang:2012mp}\\
  \hline
0.27	&77.0&	14. & \citep{Simon:2004tf}\\
  \hline
0.28	& 88.8&	36.6& \citep{Zhang:2012mp}\\
  \hline
0.35	& 76.3 & 5.6 & \citep{Chuang:2011fy}\\
  \hline
0.352&	83.0&	14.0 & \citep{Moresco:2012jh}\\
  \hline
0.4	&95.0	& 17.0 & \citep{Simon:2004tf}\\
  \hline
0.48	& 97.0&	62.0& \citep{Stern:2009ep}\\
  \hline
0.593	& 104.0	& 13.0 & \citep{Moresco:2012jh} \\
  \hline
0.68&	92.0&	8.0 & \citep{Moresco:2012jh}\\
  \hline
0.781	& 105.0 & 12. & \citep{Moresco:2012jh}\\
  \hline
0.875&	125.0&	17.0 & \citep{Moresco:2012jh} \\
  \hline
0.88	&90.0	&40.0 & \citep{Stern:2009ep}\\
  \hline
0.9	&117.0	&23.0 & \citep{Simon:2004tf}\\
  \hline
1.037&	154.0&	20.0 & \citep{Moresco:2012jh}\\
  \hline
1.3	&168.0&	17.0 & \citep{Simon:2004tf}\\
  \hline
1.43	& 177.0	& 18.0 & \citep{Simon:2004tf}\\
  \hline
1.53&	140.0&	14.0 & \citep{Simon:2004tf}\\
  \hline
1.75	& 202.0	&40.0 & \citep{Simon:2004tf}\\
  \hline
2.3&	224.0&	8.0 & \citep{Busca:2012bu} \\
 \hline \hline
\end{tabular}\label{tabh}
\end{table}

Now we combine the above cosmological observations, using a joint likelihood analysis, in order to put even more stringent constraints on the free parameter space, according to:
\begin{eqnarray}\label{eq:like-tot}
 {\cal L}_{\rm tot}({\bf p})={\cal L}_{\rm sn} \times {\cal L}_{\rm bao} \times {\cal L}_{\rm cmb} \times {\cal L}_{\rm h} \times
 {\cal L}_{\rm bbn}\;,
\end{eqnarray}

or in chi-square form
\begin{eqnarray}\label{eq:like-tot_chi}
 \chi^2_{\rm tot}({\bf p})=\chi^2_{\rm sn}+\chi^2_{\rm bao}+\chi^2_{\rm cmb}+\chi^2_{\rm h}+\chi^2_{\rm bbn}\;.
\end{eqnarray}
with the likelihood estimator defined as $ {\cal L}_{\rm j}=\exp(-\frac{\chi^2_{\rm j}}{2})$. We maximize the relevant joint likelihood function (or minimize the total chi-square function) to find the best value of free parameters. To obtain best results, we employ a Metropolis Markov Chain Monte Carlo (MCMC) procedure.

In the Bayesian framework we can update our beliefs iteratively in real time as data comes in\citep{Geyer2011}. It works as follows: we have a prior belief about the value of a parameter and some of observational data. We can update our beliefs by calculating the posterior distribution in the first loop of the chain of MCMC. In the next loop, our posterior becomes the new prior. We can update the new prior with the likelihood derived from the new step and again we get a new posterior. This cycle can continue indefinitely so we are continuously updating our beliefs. The amount of weight that we put on our prior versus our likelihood depends on the relative uncertainty between these two distributions. If we put more weight on our prior (prior distribution is much less spread out than the likelihood distribution), then the posterior resembles the prior much more than the likelihood. This approach is useful when something is wrong with the data collection process\citep{Hastings:1970aa,Trotta:2008qt}. But in our case which we have enough confirmed data points, we can choose to widen the prior distribution in relation to the likelihood. Therefore we allow the priors on the parameters to be wide enough. In the Table \ref{tab:inipar} we present the initial values which we selected for the free parameters in the first chain of our MCMC.

\begin{table}
 \centering
 \caption{The initial values of the free parameters used in the first chain of MCMC. These values for ODE1 and ODE2 obtained from \citep{Linder:2005dw} and \citep{Ma:2011nc} }
\begin{tabular}{c  c  c  c }
\hline \hline
 parameter  & ODE1 & ODE2 & $\Lambda$CDM \\
 \hline
  $\Omega_{\rm DM0}$ &  0.25 & 0.25 & 0.25 \\
 \hline
$\Omega_{\rm b0}$ & 0.04 & 0.04 & 0.04 \\
 \hline 
 $h$   &  0.65 & 0.65 & 0.65 \\
 \hline
$w_0$ &  -0.9 & -1.061 & -- \\
 \hline 
$A$ & 0.15  & 0.041 & -- \\
 \hline 
$B$ &  1.0 & 1.0 & -- \\
  \hline \hline
\end{tabular}\label{tab:inipar}
\end{table}

\begin{figure*} 
	\centering
	\includegraphics[width=13cm]{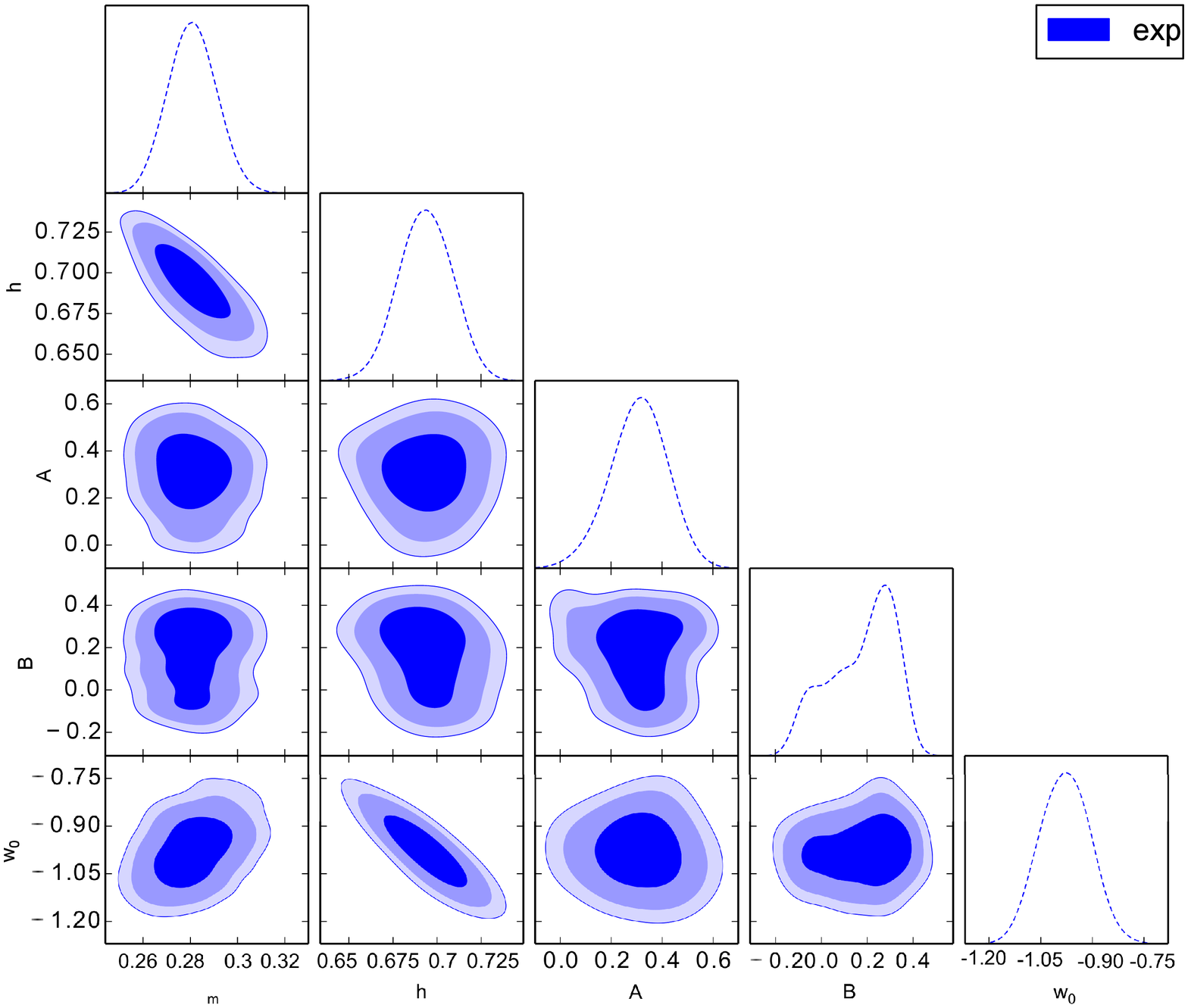}
	\includegraphics[width=13cm]{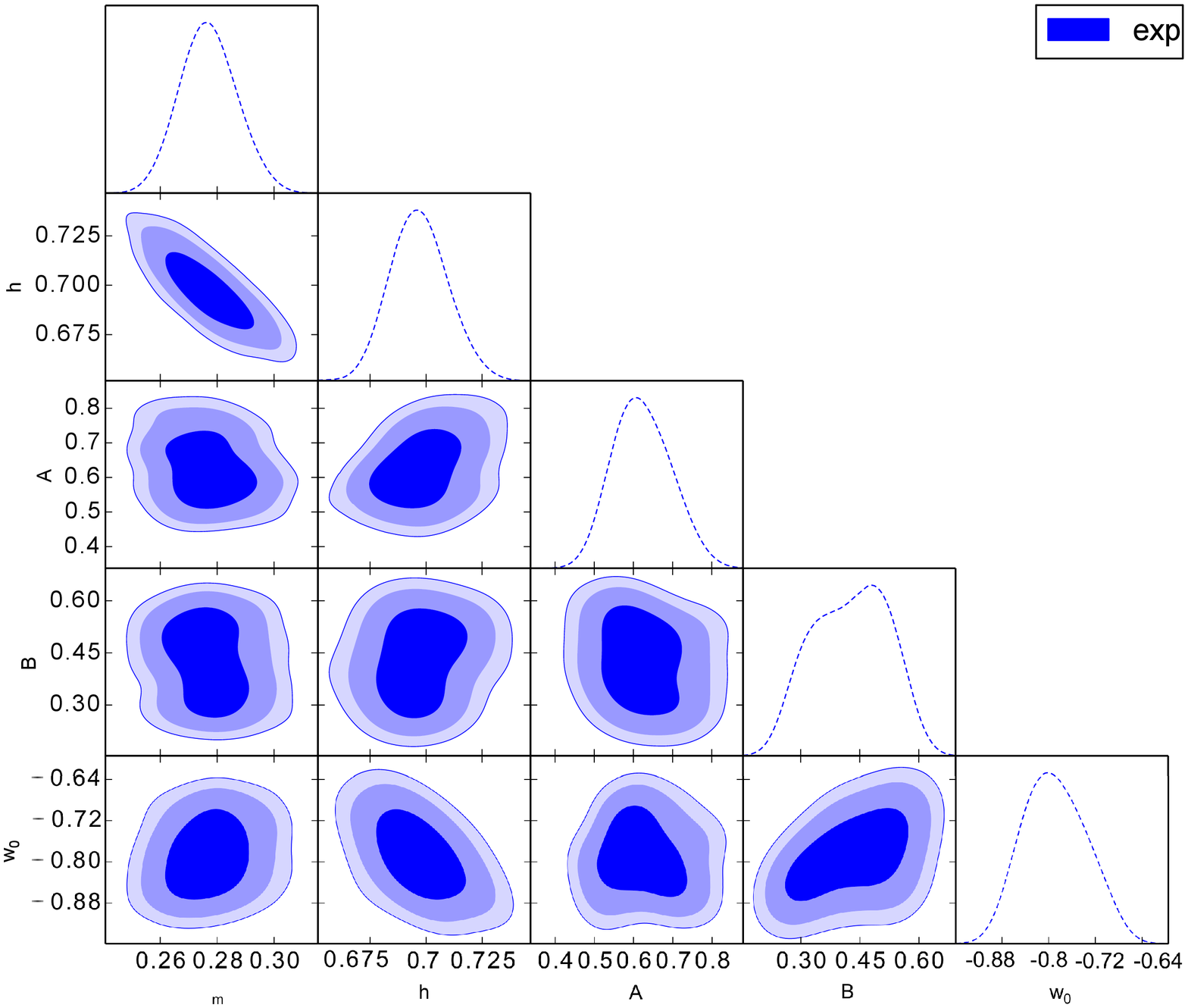}
	\caption{The $1\sigma$, $2\sigma$ and $3\sigma$ confidence contours 
for various free parameters of models using the latest background (expansion) data. The results for ODE1 and ODE2 presented in the upper and lower panels respectively.}
	\label{fig:contour}
\end{figure*}

\begin{figure} 
	\centering
	\includegraphics[width=8cm]{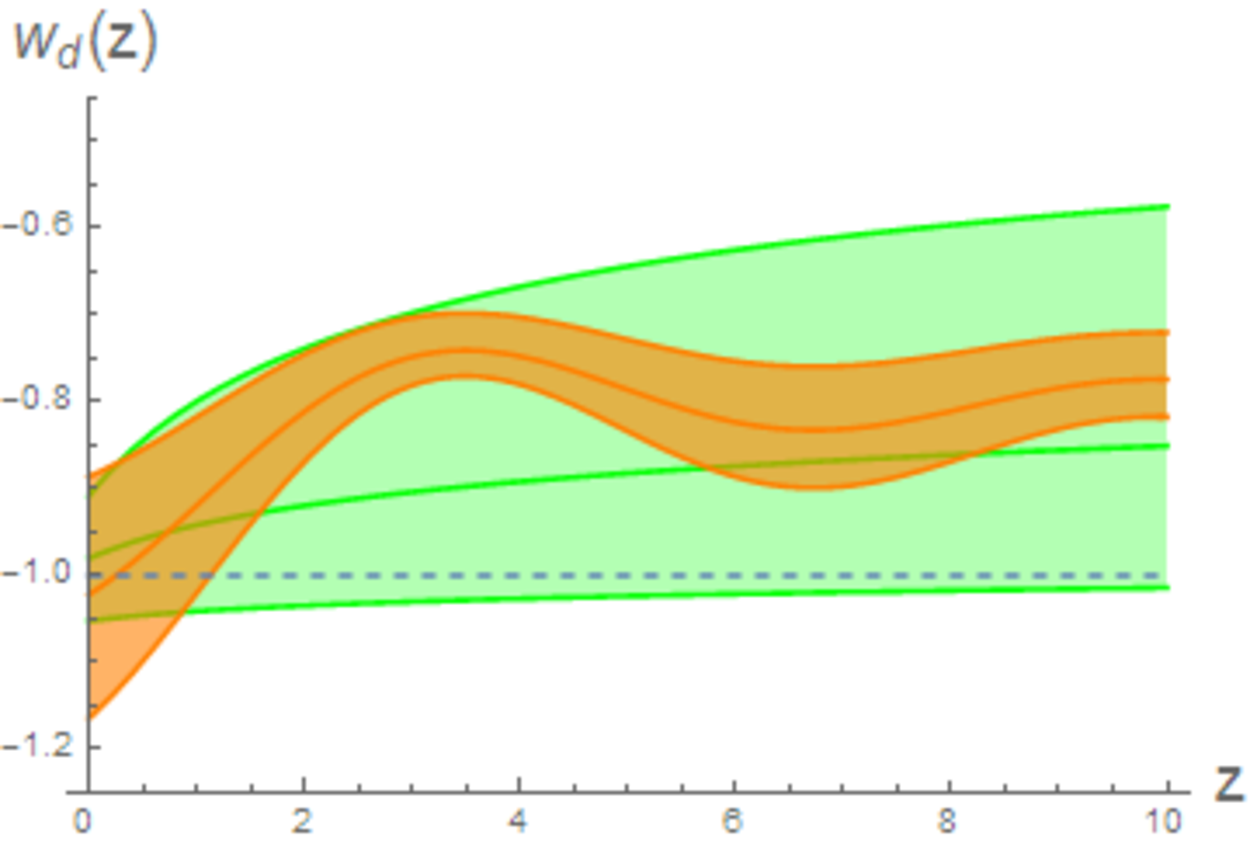}
	\includegraphics[width=8cm]{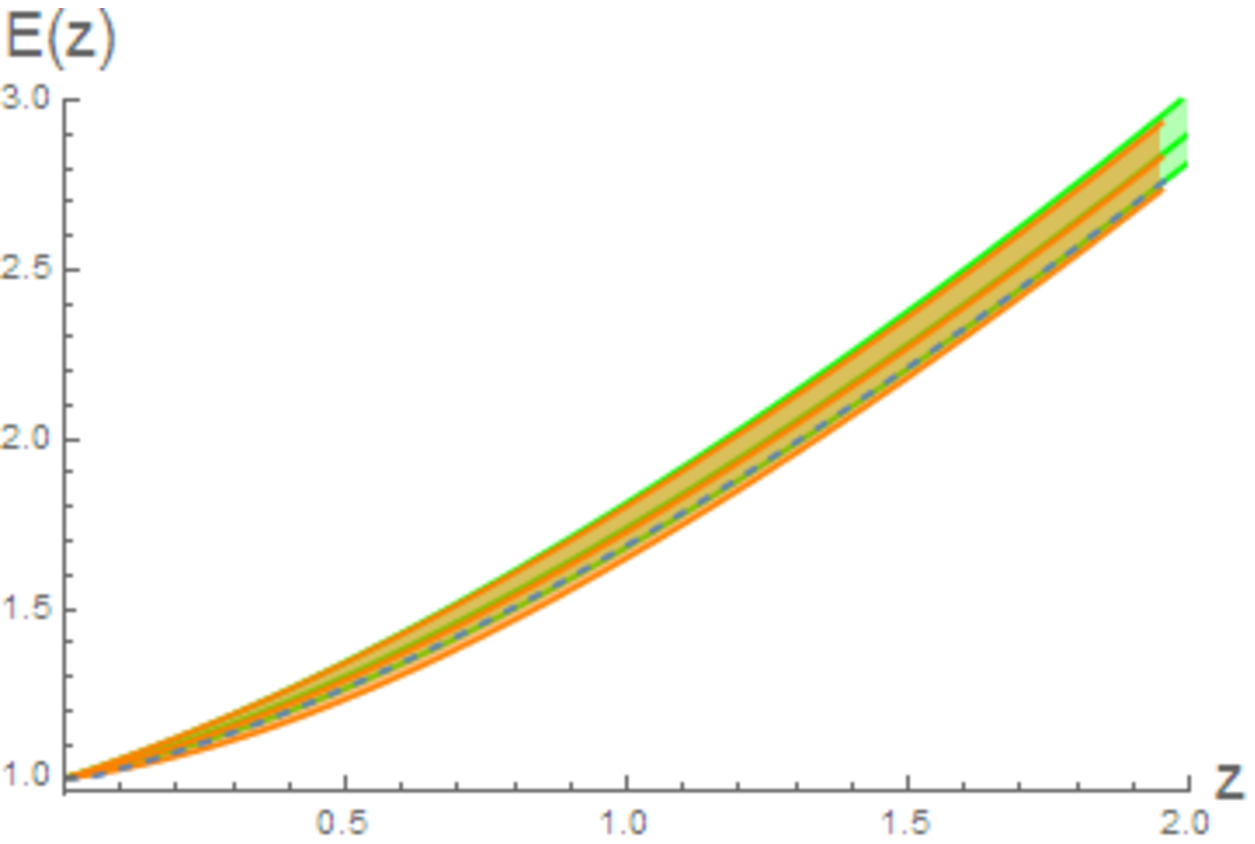}
	\caption{The evolution of $w_{\rm d}(z)$ (upper panel) and $ E(z)$ (bottom panel) for different ODE models. The green area shows the $1\sigma$ confidence level of ODE1 and orange area is the $1\sigma$  confidence level of ODE2. For comparison we showed the related results for concordance $\Lambda$CDM model with dashed line.}
	\label{fig:BG}
\end{figure}

In order to obtain the reliable values of the maximum likelihood, in our analysis we run more than $150000$ chains for each of the cases. These long chains can guarantee the reliability of the maximum likelihood value and also the best values which we find for free parameters. So, after running these long chains, we can claim that our estimating values for ${\cal L}_{\rm max}$ or $\chi^2_{\rm min}$ are also in the high level of confidence\citep{Hastings:1970aa,Geyer2011,Trotta:2017wnx}. For more details on this method we refer the reader to\citep{Mehrabi:2015hva,Malekjani:2016edh,Trotta:2017wnx,Rezaei:2017yyj,Malekjani:2018qcz}. One can see the results of our statistical analysis in Tables (\ref{tab:best}) and (\ref{tab:bestfit}) respectively. In our analysis free parameters of the model (the statistical vector ${\bf p}$ in Eqs.(\ref{eq:like-tot} \& \ref{eq:like-tot_chi})) include $\{\Omega_{\rm DM0}, \Omega_{\rm b0}, h, w_0, A, B\}$, where  $h=H_{0}/100$ and the energy density of radiation is fixed to $\Omega_{\rm r0}=2.469\times 10^{-5}h^{-2}(1.6903)$ \citep{Hinshaw:2012aka}. Additionally, in order to test the statistical performance of our models and to compare them with concordance $\Lambda$CDM model, we utilize the well known information criteria, namely  ${\rm AIC}$ which defined as\citep{Akaike:1974} 
\begin{eqnarray}
{\rm AIC} = -2 \ln {\cal L}_{\rm max}+2k\;,
\end{eqnarray}

and BIC which defined as\citep{Schwarz:1974}  
\begin{eqnarray}
{\rm BIC} = -2 \ln {\cal L}_{\rm max}+k\ln N\;.
\end{eqnarray}
where $k$ is the number of free parameters considered in MCMC procedure and $N$ is the overall number of data points. In the case of ODE models considered in this work we have $k=6$ and $N=615$ at the background level which we use only  expansion data. For $\Lambda$CDM we have $k=3$. Although ODE models in this study provide low $\chi^2_{\rm min}$ values with respect to those of $\Lambda$CDM, but due to more number of free parameters we use ${\rm AIC}$ and ${\rm BIC}$ values.
An important parameter associated with the ${\rm AIC}$ is $\Delta {\rm AIC}={\rm AIC}-{\rm AIC}_{\rm \Lambda}$ which can be used to compare models [see Table (\ref{tab:best})]. 
ODE models in this study provide relatively high $\Delta {\rm AIC}$ values with respect to $\Lambda$CDM ( $\Delta {\rm AIC}>4$), and therefore have considerably less support \citep{Burnham2002}. In the case of ${\rm BIC}$, ODE models provide $\Delta {\rm BIC}>10$, which represent very strong evidence against the model with the higher ${\rm BIC}$ value. From these, we notice a tension between ${\rm AIC}$ and ${\rm BIC}$ results, while ${\rm AIC}$ indicates there is "weak evidence against" ODE models, ${\rm BIC}$ indicates that there is "strong evidence against" ODE models. This is due to the fact that ${\rm BIC}$ strongly penalizes models with a larger number of parameters \citep{Liddle:2004nh}
In order to visualize the solution space of the model parameters, we plot the $1\sigma$, $2\sigma$ and $3\sigma$ confidence levels for different parameter pairs of oscillating DE models in Fig.(\ref{fig:contour}). Using the best fit model parameters [see Table (\ref{tab:bestfit})] 
in Fig. (\ref{fig:BG}) we plot the evolution of $w_{\rm d}(z)$ (upper panel) and $ E(z)$ (bottom panel). 
Although a small region from $1\sigma$ level of  the EoS parameter of ODE1 (green area in upper panel of Fig. (\ref{fig:BG})) is in the phantom regime, but major part of it (especially the EoS parameter of ODE1, based on best fit parameters) remain in the quintessence regime ($-1<w_{\rm d}<-1/3$). In this model because of small value of $B$, the frequency of oscillation, we can not observe the oscillatory behavior of EoS in low redshifts. On the other hand in the case of ODE2, the oscillatory behavior of EoS easily can be seen, because the best value of $B$ for this model is greater than ODE1 ones. In the recent case, $1\sigma$ region of $w_{\rm d}$ (orange area in upper panel of Fig. (\ref{fig:BG})), for $z\gtrsim1$ is in the quintessence region, while it enters in the phantom regime at relatively low redshifts.
In the bottom panel of Fig. (\ref{fig:BG}), we see the redshift evolution of $1\sigma$ region of Hubble parameter $E$ for ODE models in comparison with that of the usual $\Lambda$ cosmology (dashed line). we observe that the expansion rate of the universe in both of ODE models (based on best fit values) is larger than that of the $\Lambda$CDM. However, $E_{\Lambda}$ remains in the  $1\sigma$ region of Hubble parameters of  ODE models. Between these oscillating models, ODE1 (green) has grater $E$ value and experiences bigger expansion rate.

\begin{table}
 \centering
 \caption{The statistical results for the different ODE models. These results obtained from the background data.
The results of $\Lambda$CDM model are presented for comparison.}
\begin{tabular}{c  c  c c c c c  }
\hline \hline
 Model  & $k$ &$\chi^2_{\rm min}$ &${\rm AIC}$&$\Delta {\rm AIC}$&${\rm BIC}$&$\Delta {\rm BIC}$ \\
 \hline
ODE1  & $6$ & $563.2$  & $575.2$ & $4.8$ & 601.6 & 18.0 \\
 \hline
ODE2 & $6$ & $562.5$  & $574.5$ & $4.1$& 600.9 & 17.3 \\
 \hline 
 $\Lambda$CDM  & $3$ & $564.4$  & $570.4$ & $0$& 583.6 & 0 \\
  \hline \hline
\end{tabular}\label{tab:best}
\end{table}

\begin{table}
 \centering
 \caption{The best value of free parameters for different ODE models obtained from expansion data.
}
\begin{tabular}{c  c  c c }
\hline \hline
 Model  & ODE1 & ODE2 & $\Lambda$CDM\\
\hline 
$\Omega_{\rm m}^{(0)}$ & $0.280\pm 0.011$ & $0.276^{+0.011}_{-0.010}$ & $0.276\pm 0.0084$\\
\hline
$ h $ & $0.696\pm 0.014$ & $0.696^{+0.013}_{-0.014}$ & $0.705\pm 0.018$\\
\hline
$ A $ & $0.313^{+0.13}_{-0.11}$ & $0.619^{+0.073}_{-0.080}$ & $--$\\
\hline
$ B $ & $0.178^{+0.18}_{-0.11}$ & $0.429^{+0.11}_{-0.10}$ & $--$\\
\hline
 $ w_0$ & $-0.982\pm 0.071$ & $-0.788^{+0.05}_{-0.06}$ & $--$\\
\hline
 $ w_{\rm d}(z=0) $ & $-0.982$ & $-1.012$ & $-1.0$\\
\hline
 $ \Omega_{\rm d}(z=0) $ & $0.72$ & $0.724$ & $0.724$\\
\hline \hline
\end{tabular}\label{tab:bestfit}
\end{table}

\section{Growth of perturbations in oscillating DE cosmologies}\label{growth}

\begin{table*}
	\centering
	\caption{The best value of free parameters for homogeneous and clustered ODE models obtained from growth rate data.
}
	\resizebox{\textwidth}{!}{%
		\begin{tabular}{c  c  c c c c c}
			\hline \hline
			Model  & $\Omega_{\rm m}^{(0)}$ & $ h $  & $ A $ & $ B $ & $ w_{0} $ & $ \sigma_8 $  \\
			\hline 
			ODE1 (homogeneous) & $0.445^{+0.057 ~}_{-0.060 ~}$ & $0.7067^{+0.009 ~}_{-0.009 ~}$ & $0.270^{+0.056}_{-0.054}$ &  $0.577^{+0.067}_{-0.058}$ &  $-1.025^{+0.029 ~}_{-0.028 ~}$  &  $0.892^{+0.086 ~}_{-0.086 ~}$ \\
			\hline
			ODE1 (clustered)& $0.437^{+0.055 ~}_{-0.062 ~}$ &  $0.7041^{+0.00941 ~}_{-0.0090 ~}$ & $0.262^{+0.057 ~}_{-0.057 ~}$ &  $0.560^{+0.067}_{-0.059}$ &  $-1.037^{+0.024 ~}_{-0.026 ~}$ & $0.871^{+0.086 ~}_{-0.086 ~}$ \\
			\hline
			ODE2 (homogeneous) & $0.328^{+0.056 ~}_{-0.059 ~}$ & $0.7035^{+0.0074 ~}_{-0.0077 ~}$ & $0.605^{+0.054}_{-0.061}$ & $0.430^{+0.067}_{-0.056}$ & $-0.818^{+0.0276 ~}_{-0.0279 ~}$  &  $0.842^{+0.086 ~}_{-0.086 ~}$ \\
			\hline
			ODE2 (clustered) & $0.328^{+0.055 ~}_{-0.062 ~}$ & $0.7076^{+0.0090 ~}_{-0.0090 ~}$ & $0.597^{+0.049}_{-0.060}$ & $0.434^{+0.052}_{-0.050}$ & $-0.8095^{+0.028}_{-0.028}$  & $0.841^{+0.082 ~}_{-0.082 ~}$ \\
			\hline
			$\Lambda$CDM & $0.281^{+0.009 ~}_{-0.009 ~}$ & $0.926^{+0.0090 ~}_{-0.0090 ~}$ & $--$ & $--$ & $--$  & $0.787^{+0.029 ~}_{-0.029 ~}$ \\

			\hline \hline
		\end{tabular}\label{tab:bestfitgronly}
	}
\end{table*}

\begin{figure*} 
	\centering
	\includegraphics[width=8cm]{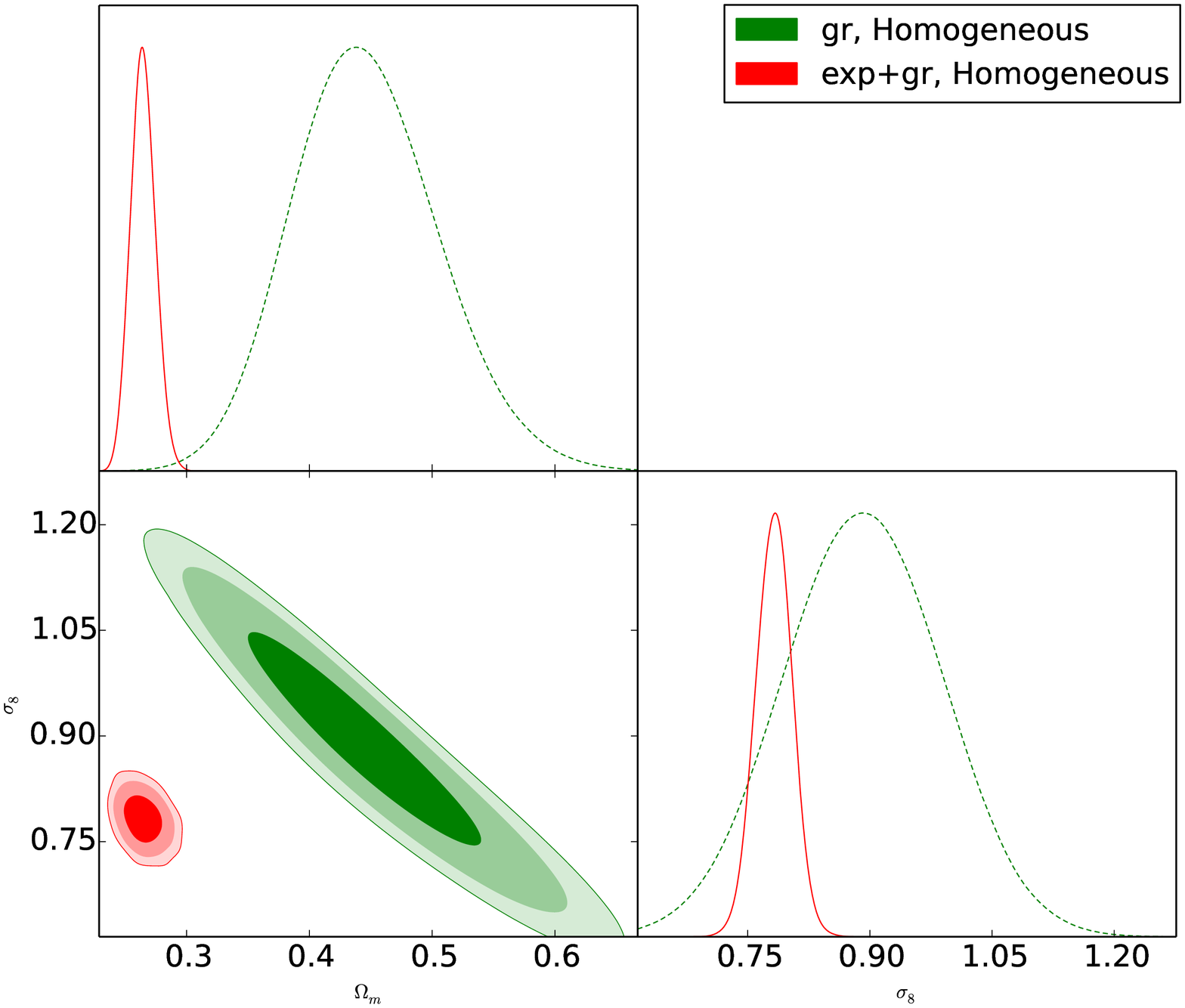}
	\includegraphics[width=8cm]{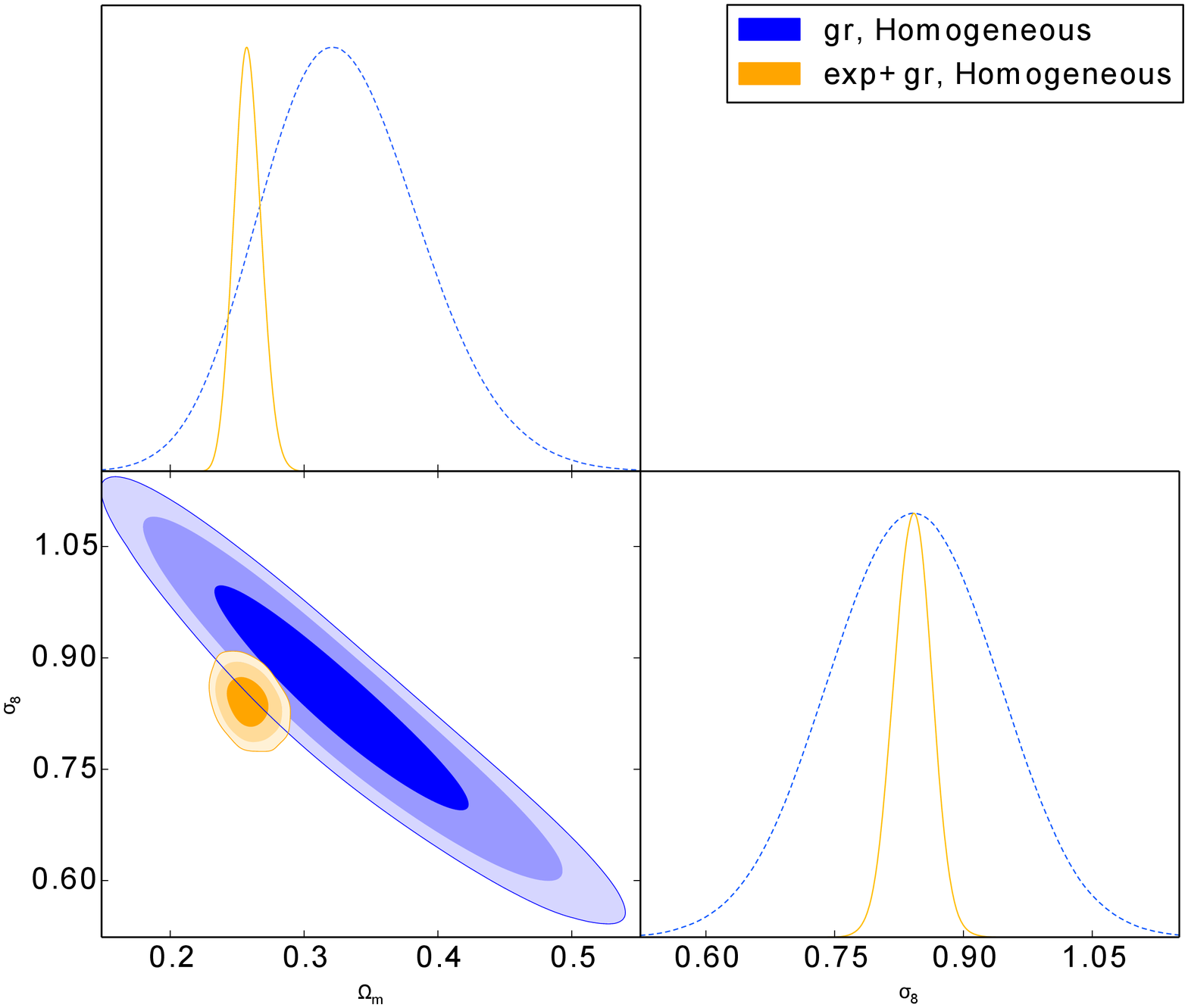}
	\includegraphics[width=8cm]{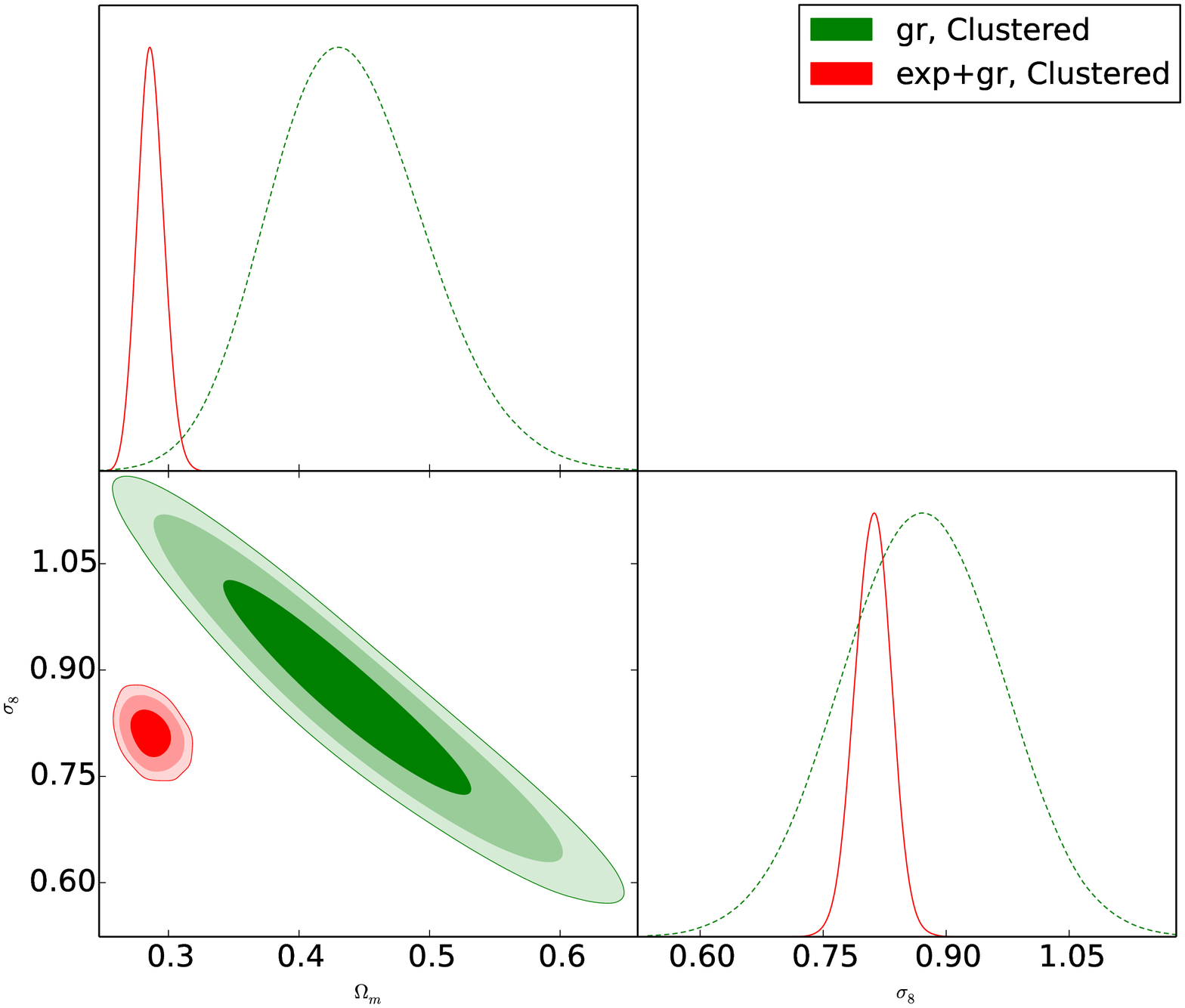}
	\includegraphics[width=8cm]{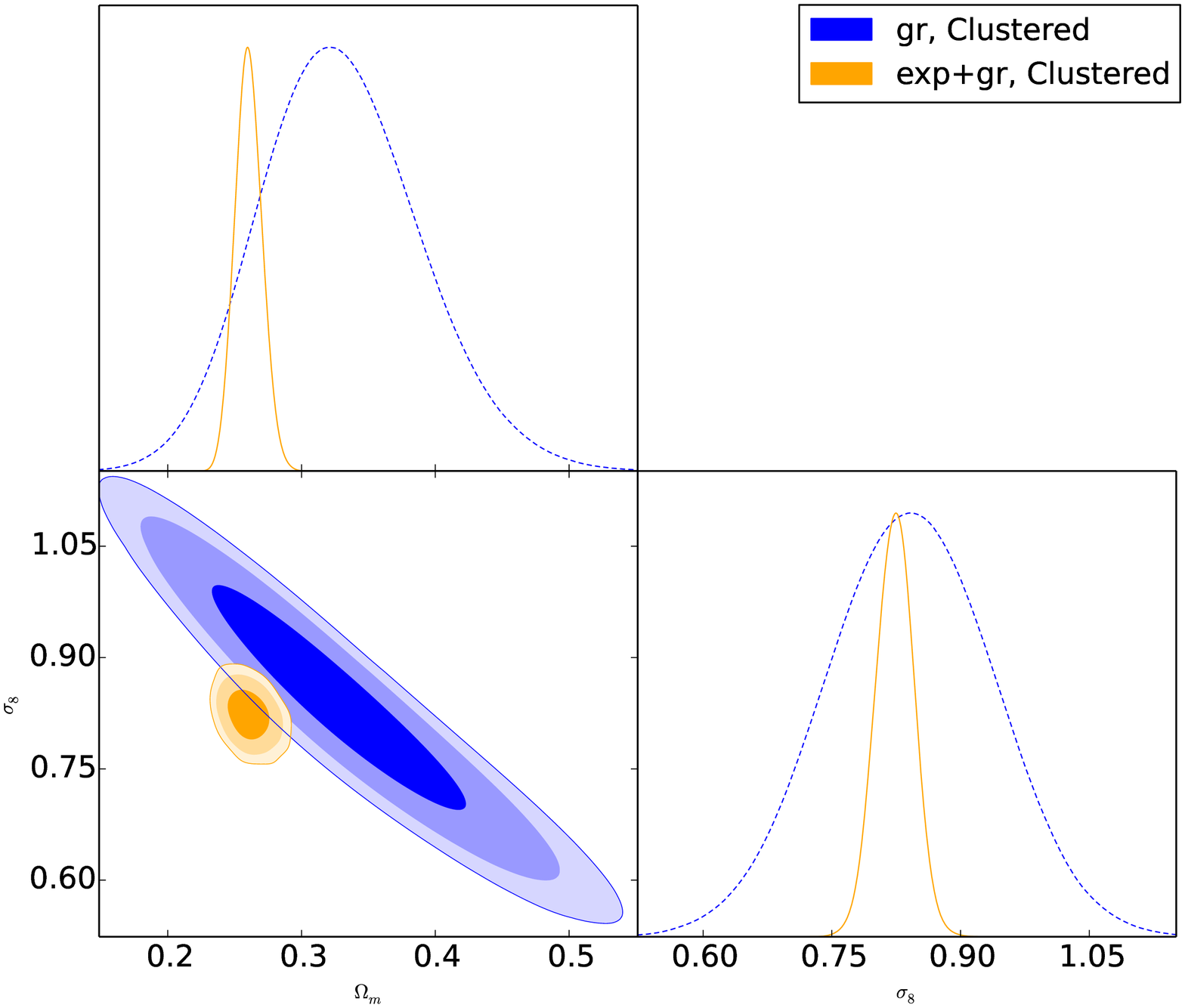}
	\caption{ The $1\sigma$, $2\sigma$ and $3\sigma$  likelihood contours 
for $\sigma_8-\Omega_{\rm m}$ plane obtained from solely growth rate data (gr) and combined data (exp+gr) for homogeneous and clustered ODE models. In the upper left (upper right) panel one can see the results for homogeneous ODE1 (ODE2). In the lower left (lower right) panel we present the results for clustered ODE1 (ODE2) .}
	\label{fig:contour2}
\end{figure*}

\begin{table*}
	\centering
	\caption{The statistical results for homogeneous (clustered) ODE models. These results obtained from combined data (exp+gr). The results of $\Lambda$CDM model presented for comparison.}
	\begin{tabular}{c  c  c c c c c}
		\hline \hline
 Model  & $k$ &$\chi^2_{\rm min}$ &${\rm AIC}$&$\Delta {\rm AIC}$ & ${\rm BIC}$&$\Delta {\rm BIC}$ \\
 \hline
ODE1  & 7 & 579.8 (580.2)  & 593.8 (594.2) & 9.7 (10.1) & 624.9 (625.3) & 22.9 (23.3)\\
 \hline
ODE2 & 7 & 574.1 (574.3)  & 588.1 (588.3) & 4.0 (4.2) & 619.2 (619.4) & 17.2 (17.4)\\
 \hline 
 $\Lambda$CDM  & 4 & 576.1  & 584.1 & 0 & 602.0 & 0 \\
  \hline \hline
\end{tabular}\label{tab:best2}
\end{table*}

\begin{table*}
	\centering
	\caption{The best values of free parameters for both homogeneous and clustered ODE models using the combined data (exp+gr).
}
	\resizebox{\textwidth}{!}{%
		\begin{tabular}{c  c  c c c c c}
			\hline \hline
			Model  & $\Omega_{\rm m}^{(0)}$ & $ h $  & $ A $ & $ B $ & $ w_{0} $ & $ \sigma_8 $  \\
			\hline 
			ODE1 (homogeneous) & $0.265^{+0.011~}_{-0.011~}$ & $0.7093^{+0.0091 ~}_{-0.0091 ~}$ & $0.271^{+0.054}_{-0.050}$ &  $0.572^{+0.071}_{-0.058}$ &  $-1.0276^{+0.024 ~}_{-0.024 ~}$ & $0.782 ^{+0.021 ~}_{-0.021 ~}$\\
			\hline
			ODE1 (clustered)& $0.286^{+0.010 ~}_{-0.010 ~}$ & $0.7074^{+0.0081 ~}_{-0.0081 ~}$ & $0.265^{+0.051 ~}_{-0.051 ~}$ &  $0.563^{+0.065}_{-0.059}$ &  $-1.034^{+0.0247 ~}_{-0.0246 ~}$ & $0.811 ^{+0.022 ~}_{-0.022 ~}$\\
			\hline
			ODE2 (homogeneous) & $0.258^{+0.0099 ~}_{-0.0099 ~}$ & $0.7009^{+0.009 ~}_{-0.009 ~}$ & $0.6035^{+0.054}_{-0.058}$ & $0.432^{+0.067}_{-0.051}$ & $-0.818^{+0.028 ~}_{-0.028 ~}$ & $0.841^{+0.022 ~}_{-0.022 ~}$\\
			\hline
			ODE2 (clustered) & $0.260^{+0.01 ~}_{-0.01 ~}$ & $0.7018^{+0.0090 ~}_{-0.0090 ~}$ & $0.594^{+0.053}_{-0.059}$ & $0.435^{+0.052}_{-0.050}$ & $-0.809^{+0.026}_{-0.026}$  & $0.823^{+0.022 ~}_{-0.022 ~}$\\
				\hline
			$\Lambda$CDM & $0.2752^{+0.009 ~}_{-0.009 ~}$ & $0.7039^{+0.0067 ~}_{-0.0067 ~}$ & $--$ & $--$  & $--$  & $0.7679^{+0.011 ~}_{-0.011 ~}$ \\
			\hline \hline
		\end{tabular}\label{tab:bestfit2}
	}
\end{table*}

Dark energy affects large scale structures through three different mechanisms. In first mechanism, dark energy increases the expansion rate of the universe, so it suppresses the formation of structures. Furthermore, as dark energy becomes the dominating component of the Universe, it slows down the growth of large scale overdensities, and the process of gravitational structure formation will reduce at scales comparable to the Hubble distance. This two mechanisms affect the formation of structures through changes in the Hubble expansion rate. But in third manner dark energy can affect the rate of structure formation directly. If dark energy can fluctuate, not only it feels the gravitational pull of dark matter structures, but it tends to form structures itself. Thus, when we want to investigate the nature of this exotic fluid, it will be more useful to study its effects on the formation of structures in the universe. In order to study the effects of oscillating dark energy on the growth of matter fluctuations in linear regime, we compare our model forecasts for growth rate of structures, with related observational data. with the aim to distinguish oscillating dark energy from cosmological constant, we choose two different condition for dark energy, clustering DE and homogeneous DE. In the first condition, $c_{\rm eff}$, the effective sound speed of dark energy is negligible and thus, dark energy can be clustered. In this case, perturbations of DE can grow same as matter perturbations \citep[see also][]{Abramo:2008ip,Batista:2013oca,Batista:2014uoa}. On the other hand, one can choose $c_{\rm eff}=1$ and therefore dark energy remain homogeneous. Following the procedure have been used for other dark energy models in literatures \citep{Abramo:2008ip,Mehrabi:2015kta,Malekjani:2016edh,Rezaei:2017yyj,Rezaei:2017hon}, we compute the evolution of matter and dark energy perturbations ($\delta_{\rm m}$ and $\delta_{\rm d}$)  by  

\begin{eqnarray}
 &&\dot{\delta_{\rm m}}+\frac{\theta_{\rm m}}{a}=0\;,\label{grwoth1}\\
&& \dot{\delta_{\rm d}}+(1+w_{\rm d})\frac{\theta_{\rm d}}{a}+3H(c_{\rm eff} ^2 -w_{\rm de})\delta_{\rm d}=0\;,\label{grwoth2}\\
&& \dot{\theta_{\rm m}}+H \theta_{\rm m} - \frac{k^2 \phi}{a}=0\;,\label{grwoth3}\\
 &&\dot{\theta_{\rm d}}+H \theta_{\rm d} - \frac{k^2 c_{\rm eff} ^2 \theta_{\rm d}}{(1+w_{\rm d})a} - \frac{k^2 \phi}{a} =0\;,\label{grwoth4}
\end{eqnarray}

where dot means derivative with respect to cosmic time $t$, and $k$ and $c_{\rm eff}$ are the wavenumber and effective sound speed of perturbations \citep{Abramo:2008ip,Batista:2013oca,Batista:2014uoa}.
Now, using the Poisson equation in sub-Hubble scales and Combining it with the Eqs.(\ref{grwoth1}-\ref{grwoth4}), we have

 \begin{eqnarray}
\label{Possnew}
 -\frac{k^2}{a^2}\phi=\frac{3}{2} H^2[\Omega_{\rm m} \delta_{\rm m} + (1+3 c_{\rm eff} ^2)\Omega_{\rm d} \delta_{\rm d}]\;,
\end{eqnarray}
 with Eqs. (\ref{grwoth3} \& \ref{grwoth4}), eliminating $\theta_{\rm m}$ and $\theta_{\rm d}$ and changing the time derivative to scale factor $a$, we have \citep{Malekjani:2016edh,Rezaei:2017yyj}
\begin{eqnarray}
 \delta_{\rm m}''+\frac{3}{2 a}(1-w_{\rm d} \Omega_{\rm d}) \delta_{\rm m}'=\frac{3}{2a^2}[\Omega_{\rm m} \delta_{\rm m} +\Omega_{\rm d}  (1+3 c_{\rm eff} ^2)\delta_{\rm d}]\;,\label{growth5}\\
 \delta_{\rm d}''+A \delta_{\rm d}'+B \delta_{\rm d}=\frac{3}{2a^2}(1+w_{\rm d})[\Omega_{\rm m} \delta_{\rm m} +\Omega_{\rm d}  (1+3 c_{\rm eff} ^2)\delta_{\rm d}]\;\label{growth6}.
 \end{eqnarray}
where prime denotes derivative with respect to scale factor and the coefficients $A$ and $B$ have the form
\begin{eqnarray}
&&A=\frac{1}{a}[-3 w_{\rm d} -\frac{a w_{\rm d}'}{1+w_{\rm d}}+\frac{3}{2}(1-w_{\rm d} \Omega_{\rm d})],\nonumber \\  
&&B=\frac{1}{a^2}[-a w_{\rm d}' +\frac{a w_{\rm d}' w_{\rm d}}{1+w_{\rm d}}-\frac{1}{2}w_{\rm d}(1-3 w_{\rm d} \Omega_{\rm d})].
\end{eqnarray}

We solve the system of Eqs.(\ref{growth5} \& \ref{growth6}) numerically. Concerning
the initial conditions, we impose the following restrictions: at $a_i = 0.0005$, we use $\delta_{\rm mi}=\delta_{\rm m}(a_i)= 5 \times 10^{-5}$. Additionally, we insert the initial conditions as follows\citep[see also][]{Batista:2013oca,Malekjani:2016edh}
 \begin{eqnarray}\label{initialcondition}
&& \delta_{\rm mi}'=\frac{\delta_{\rm mi}}{a_{\rm i}}\;,\nonumber \\  
&& \delta_{\rm di}=\frac{1+w_{\rm di}}{1-3w_{\rm di}}\delta_{\rm mi}\;,\nonumber \\  
&&\delta_{\rm di}'=\frac{4 w_{\rm di}'}{(1-3w_{\rm di})^2}\delta_{\rm mi}+\frac{1+w_{\rm di}}{1-3w_{\rm di}}\delta_{\rm mi}'\;,
\end{eqnarray}
Using these initial values one can be confident that matter perturbations always remain in the linear regime.
After  finding the evolution of fluctuations ($\delta_{\rm m}, \delta_{\rm d}$) we can calculate the growth rate of large scale structures in the presence of oscillating dark energy models considered in this work. The growth rate function can be written as
\begin{eqnarray}\label{f(z)}
f(z)=\frac{d\ln{\delta_{\rm m}}}{d\ln{a}}\;.
\end{eqnarray}

The value of $\sigma_8(z)$, the matter fluctuation amplitude on scales of $8Mpch^{-1}$, also can be written as.
\begin{eqnarray}\label{s8(z)}
\sigma_8(z)=\frac{\delta_{\rm m}(z)}{\delta_{\rm m}(z=0)}\sigma_8(z=0)\;.
\end{eqnarray}
By multiplication of these two functions, one can calculate the value of $f(z)\sigma_8(z)$ to compare it with $f \sigma_8$ observational data.

The growth rate data set were obtained from redshift space distortions from different galaxy surveys.  These data points, when measured using redshift-space distortions, are degenerate with the Alcock-Paczynski (AP) effect \citep{Alcock:1979mp}. Assuming an incorrect cosmological model for the coordinate transformation from redshift space to comoving space leads to residual geometric distortions known as the redshift space distortions (RSD). These distortions are induced by the fact that measured distances along and perpendicular to the line of sight are fundamentally different. Measuring the ratio of galaxy clustering in the radial and transverse directions provides a probe of AP effect. However, given high precision clustering measurements over a wide range of scales, this degeneracy can be broken since RSD and AP have different scale-dependences \citep{Reid:2012sw}.  In order to overcome the RSD problem, several methods have been used in the literature to apply the AP test to the large scale structure \citep{Song:2008qt,Samushia:2011cs,Blake:2012pj,Hudson:2012gt,Blake:2013nif,Chuang:2013wga,Howlett:2014opa,Feix:2015dla,Huterer:2016uyq}. Since we obtain our growth rate data from the above references, therefore we can claim that the AP effect has been considered for the growth rate data which we used in our analysis. Nowadays there are more than $30$ data points of recent  $f\sigma_8$ measurements from different surveys. The information in some of these data points overlaps significantly with other data points in this collection. Some of them are updated version of previous measurements either with enhancements in the volume of the survey, during its scheduled run or with different methodologies by various groups. Therefore, the collection of these data points should not be used in its entirety. Therefore, in our analysis we use the "Gold-2017" compilation of robust and independent $f\sigma_8$ measurements from different surveys constructed by \citep{Nesseris:2017vor}. These $18$ data points and their references are shown in Tab.\ref{tabfs8data}.
\begin{table}
 \centering
 \caption{The $f\sigma_8$ data points and their references. }
\begin{tabular}{c  c  c c c }
\hline \hline
 $z$  & $f\sigma_8(z)$ & $\sigma_{f\sigma_8}$&  Reference \\
 \hline
 0.02 & 0.428 & 0.0465 &  \citep{Huterer:2016uyq}   \\
 \hline 
 0.02 & 0.398 & 0.065 &  \citep{Hudson:2012gt}   \\
 \hline 
  0.02 & 0.314 & 0.048 &  \citep{Hudson:2012gt}   \\
 \hline
 0.10 & 0.370 & 0.130 &  \citep{Feix:2015dla} \\
 \hline 
  0.15 & 0.490 & 0.145&  \citep{Howlett:2014opa} \\
 \hline
 0.17 &	0.510 &0.060 &  \citep{Song:2008qt} \\
 \hline
 0.18 & 0.360 & 0.090 &  \citep{Blake:2013nif} \\
 \hline
 0.38 & 0.440 & 0.060 &  \citep{Blake:2013nif} \\
 \hline
 0.25 & 0.3512 & 0.0583&  \citep{Samushia:2011cs} \\
 \hline
 0.37& 0.4602 & 0.0378&  \citep{Samushia:2011cs} \\
 \hline
 0.32 & 0.384 & 0.095&  \citep{Sanchez:2013tga} \\
 \hline
 0.59 & 0.488 & 0.060&  \citep{Chuang:2013wga} \\
 \hline
 0.44 & 0.413 & 0.080&  \citep{Blake:2012pj} \\
 \hline
 0.60 & 0.390 & 0.063&  \citep{Blake:2012pj} \\
 \hline
 0.73 & 0.437 & 0.072&  \citep{Blake:2012pj} \\
 \hline
 0.60 & 0.550 & 0.120&  \citep{Pezzotta:2016gbo} \\
 \hline
 0.86 & 0.400 & 0.110&  \citep{Pezzotta:2016gbo} \\
 \hline
 1.40 & 0.482 & 0.116&  \citep{Okumura:2015lvp} \\ 
 \hline \hline
\end{tabular}\label{tabfs8data}
\end{table}

We apply these data points to put constraints on model parameters. The likelihood function becomes 

\begin{eqnarray}
 {\cal L}_{\rm tot}({\bf p})= {\cal L}_{\rm gr}\;.
\end{eqnarray}
where the statistical vector ${\bf p}$ is $\lbrace\Omega_{\rm DM0},\Omega_{\rm b0}, h, w_0, A, B, \sigma_8\rbrace$ and $\sigma_8$ is the present value of $\sigma_8(z)$. The best fit parameters and $1\sigma$ errors for each of model parameters can be seen in Table (\ref{tab:bestfitgronly}). Our results show that the growth rate data, in comparison with expansion data, can not place strong constraints on the cosmological model parameters especially on $h, \Omega_{\rm M}$ and $\sigma_8$. The $1\sigma$, $2\sigma$ and $3\sigma$ confidence regions in the $\sigma_8-\Omega_{\rm m}$ plane  (green and blue contours related to ODE1 and ODE2 respectively) in Fig. \ref{fig:contour2} can prove this claim very well. To strengthen our constraint, in this step we use growth data in addition to the expansion data, to perform an overall likelihood analysis. To compute the overall likelihood function we should import the likelihood function of the growth data in Eq.(\ref{eq:like-tot}) as below
\begin{eqnarray}
 {\cal L}_{\rm tot}({\bf p})={\cal L}_{\rm sn} \times {\cal L}_{\rm bao} \times {\cal L}_{\rm cmb} \times {\cal L}_{\rm h} \times
 {\cal L}_{\rm bbn}\times {\cal L}_{\rm gr}\;.
\end{eqnarray}
Thus, the total chi-square $\chi^2_{\rm tot}$ is given by
\begin{eqnarray}
\chi^2_{\rm tot}({\bf p})=\chi^2_{\rm sn}+\chi^2_{\rm bao}+\chi^2_{\rm cmb}+\chi^2_{\rm h}+\chi^2_{\rm bbn}+\chi^{2}_{\rm gr}\;,
\end{eqnarray}

As mentioned before, the vector ${\bf p}$ contains the free parameters of the models which in this step are $\lbrace\Omega_{\rm DM0},\Omega_{\rm b0}, h, w_0, A, B, \sigma_8\rbrace$. The results of our analysis for two ODE models considered in this work, are shown in Tables (\ref{tab:best2} \& \ref{tab:bestfit2}). Also, the effects of performing overall likelihood can be seen in Fig. \ref{fig:contour2}, where adding growth data to expansion data leads to smaller confidence regions (red and orange contours). These results was obtained for both of ODE models in homogeneous and clustered DE scenarios.
Comparing the latter results with those of Section (\ref{sect:BG}), we find that putting observational constraints on the models under study using combined data (exp+gr) are practically very close to those obtained by the expansion data, in both of homogeneous and clustered ODE scenarios. Although clustering of DE changes the results of structure formation in the universe, but in comparison with observational growth data the effect of clustering is not significant. As one can see in Table \ref{tab:bestfit2}, differences between results of clustering and homogeneous approaches are negligible. Also, using the numerical results of overall likelihood analysis in Tables (\ref{tab:best2} \& \ref{tab:bestfit2}) we can not prefer one of these approaches to another one. The results of overall likelihood using combined data indicate that the $\Lambda$CDM model is the best model among models were studied in this work [see Table (\ref{tab:best2})] . Using  expansion + growth  data, the ${\rm AIC}$ test suggests weak evidence against ODE2 ($\Delta {\rm AIC} > 4$), while in the case of ODE1, $\Delta {\rm AIC} > 10$ indicates strong evidence against (essentially no support) this model. Under {\rm BIC}, there is very strong evidence against both of ODE models compared to $\Lambda$CDM. 
These results are in full agreement with those we obtain from Fig.\ref{fig:contour2}, especially in the left panels which are related to ODE1, where the posterior contours on the parameters using the growth only and combined data do not overlap even at $3\sigma$ level. We can conclude that ODE1 not only can not fit the expansion and combined data very good, but also it obtain worse results in comparison with growth data.
   
\begin{figure} 
	\centering
	\includegraphics[width=8cm]{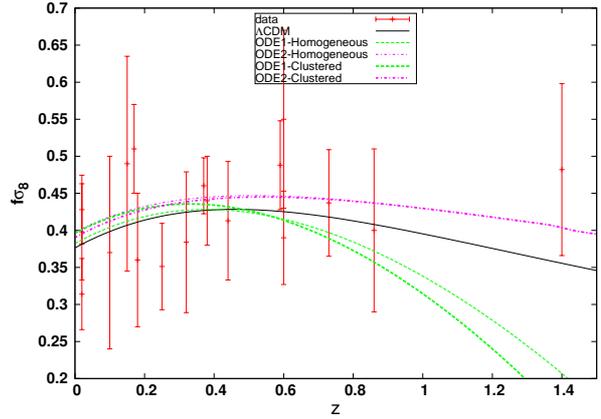}
	\caption{ Comparison between the observational data and the redshift evolution of theoretical value of $f\sigma_8 (z)$. The observational data and those errorbars showed by red lines.
		Line styles and colors for different models are shown in the legend.}
	\label{fig:fsigma8}
\end{figure}

Finally, in Fig. (\ref{fig:fsigma8}), we compare the observed $f\sigma_8(z)$ with the theoretical value of growth rate function for various ODE models. As we expected from statistical analysis, predicted growth rate in ODE1 can not fit   the observed $f\sigma_8(z)$ as well as another models. Although, at low redshifts ($z\lesssim0.4$), theoretical results are close to observational data, but at $z\gtrsim0.4$ predicted growth rate fall down and recede from observational data. The results obtained for ODE2 are more consistent with current observational
data in comparison with ODE1. However, as expected from ${\rm AIC}$ analysis (see Table \ref{tab:best2}) standard $\Lambda$CDM has the top rank in goodness of fit to data. 

\section{Conclusions}\label{conlusion}

We investigated the cosmological properties of oscillating DE models in which the EoS parameter is an oscillating function and compared how well they fit the observational data. Oscillating DE models are interesting because they are good candidates to alleviate the coincidence problem. Initially, we introduced two oscillating parameterizations for the EoS of DE and then we investigated the behavior of these ODE models at the background using the latest expansion data (SNIa, BBN, BAO, CMB and H(z)). Applying these datasets with a Markov Chain Monte Carlo (MCMC) procedure we placed constraints on the free parameters of  oscillating models. Using the well known Akaike information criteria we conclude that there is weak evidence against ODE models, while Bayesian information criteria indicated that there is very strong evidence against models under study. Using best fit values and their $1\sigma$ levels we plotted the evolution of EoS parameter and Hubble parameter for ODE models in Fig. (\ref{fig:BG}). We found that oscillatory behavior of EoS can be seen at low redshifts in ODE2. Also, the expansion rate of the universe in ODE models is bigger than that of $\Lambda$CDM. Then we studied the evolution of perturbations and  placed new constraints on the free parameters of ODE models using growth rate data. We found that the growth rate data can not place strong constraints on the ODE model parameters. Finally, we combined all of the expansion and growth data to implement an overall statistical analysis. The ${\rm AIC}$ test showed that combined data disfavor ODE1, while there is weak evidence against ODE2. The  ${\rm BIC}$ results were a bit different, from ${\rm AIC}$ ones. From ${\rm BIC}$ values we found that there is very strong evidences against ODE1 and ODE2. We did not see significant difference between homogeneous and clustered DE scenarios in our models. Combining the  ${\rm AIC}$ and  ${\rm BIC}$ results we found that among these models, $\Lambda$CDM is the best ones while the cosmological data disfavor ODE1, regard-less the status of the DE component (homogeneous or clustered). In the case of ODE2, we noted a tension between the results using ${\rm AIC}$ and  ${\rm BIC}$. Under ${\rm BIC}$ there is strong evidence against this model and we can reject it, while under ${\rm AIC}$ there is weak evidence against ODE2.

\section{Acknowledgements}
This work has been supported financially by Research Institute for Astronomy \& Astrophysics of Maragha
(RIAAM) under research project No. 1/5440-36.

\bibliographystyle{mnras}
\bibliography{ref}
\label{lastpage}

\end{document}